\documentstyle[referee]{laa}

\begin{document}
\thesaurus{07  
	   (07.09.1; 
	    07.13.1;  
	   )}

\title{On Equation of Motion for Arbitrarily Shaped Particle
under Action of Electromagnetic Radiation}
\author{J.~Kla\v{c}ka}
\institute{Institute of Astronomy,
   Faculty for Mathematics and Physics, Comenius University \\
   Mlynsk\'{a} dolina, 842~48 Bratislava, Slovak Republic}
\date{}
\maketitle

\begin{abstract}
Arguments of astronomers against equation of motion for arbitrarily
shaped particle under action of electromagnetic radiation are discussed.
Each of the arguments is commented in detail from the point of view of the
required physics. It is shown that the arguments of astronomers, including
referees in several astronomical and astrophysical journals, are unacceptable
from the physical point of view. Detail explanations should help astronomers
in better physical understanding of the equation of motion.
Relativistically covariant equation of motion for real dust particle
under the action of electromagnetic radiation is derived. The particle
is neutral in charge. Equation of motion is expressed in terms of
particle's optical properties, standardly used in optics for stationary
particles.

\keywords{relativity theory, electromagnetic radiation, cosmic dust}

\end{abstract}

\section{Introduction}
Astronomers each year publish several papers on orbital evolution of cosmic
(dust) particles which consider interaction of the particles with
electromagnetic radiation. This is taken in the form of the Poynting-Robertson
effect (P-R effect; Robertson 1937). However, P-R effect corresponds to a very
special form of interaction between the particle electromagnetic radiation
(see Eqs. (120), (122) in Kla\v{c}ka 1992a). Since real particles do not fulfill
this special form of interaction, one needs to have more general
equation of motion in disposal. This is presented in Kla\v{c}ka (2000a).
Later on, more simple derivations were presented: Kla\v{c}ka (2000b),
Kla\v{c}ka and Kocifaj (2001a), see also review paper Kla\v{c}ka (2001b).
Applications of the equation of motion can be found in Kla\v{c}ka (2000c),
Kla\v{c}ka and Kocifaj (2001a), Kla\v{c}ka and Kocifaj (2001b),
Kla\v{c}ka and Kocifaj (2002), Kocifaj and Kla\v{c}ka (2002a, 2002b).

Our experience shows that astronomers do not understand the physics of the
equation of motion. This is the reason why we have decided to present
arguments against the equation of motion and, of course, to give detail
explanations why the arguments are physically incorrect. We can say that
this paper is a continuation of the paper Kla\v{c}ka (1993).

\section{Arguments and answers}
This section presents arguments/statements of astronomers against the equation
of motion (since the year 1998) and our point of view.

\subsection{Argument/statement 1}
{\it Astronomers do not need any more general equation of motion --
they are satisfied with the Poynting-Robertson effect}.

{\it Answer:}

Scattering of light, electromagnetic radiation, on arbitrarily shaped
(dust) particles with various optical properties may significantly differ
from that required by the P-R effect: compare Kla\v{c}ka (1992a) and
Kocifaj and Kla\v{c}ka (1999). As a consequence, orbital evolution of
real particle practically always differs from that corresponding to the
P-R effect: Kocifaj et al. (2000),
Kla\v{c}ka and Kocifaj (2001a), Kla\v{c}ka and Kocifaj (2001b),
Kla\v{c}ka and Kocifaj (2002), Kocifaj and Kla\v{c}ka (2002a, 2002b).

\subsection{Argument/statement 2}
{\it Orthonormality of unit vectors used in the scattering theory holds
also in the frame of reference in which particle moves with velocity
$\vec{v}$ (also for $v \ne 0$)}.

{\it Answer:}

Although this access was used in Kla\v{c}ka (1994), Kla\v{c}ka and Kocifaj (1994)
and Kocifaj et al. (2000), it is not physically correct.
Unit vectors must be orthonormal in the proper frame of reference
(the rest frame of the particle, i. e. a moving frame relative to the
source of electromagnetic radiation). The requirement that "final
equation of motion has to be written in a covariant form" offers only
one possibility: vectors are not orthonormal in any other frame of
reference. And physical interpretation is also evident -- aberration
of light (Kla\v{c}ka 2000a, 2000b). Thus, the statement about the conservation
of normality of vectors under Lorentz transformation is nonphysical.

However, experience shows that astronomers do not accept this general
argument.
Thus, we will present a simple example which shows that access of
Kla\v{c}ka (2000a) is correct (at least for the simple example). \\

\noindent
{\bf EXAMPLE:} \\

   Let us consider a plane mirror moving (at a given moment) along
   x-axis (system S) with velocity $\vec{v} = ( v, 0, 0 )$, $v > 0$;
   the mirror is perpendicular to the x-axis (the plane of the mirror is
   parallel to the yz-plane). A beam of incident (hitting) photons is
   characterized by unit vector
   $\vec{S} ' = ( \cos \theta ', \sin \theta ', 0 )$
   in the proper frame (primed quantities) of the mirror. Reflected beam
   is described by the unit vector
   $\vec{e} ' = ( -~ \cos \theta ', \sin \theta ', 0 )$
   (in the proper frame S'). \\

   The problem is: Find equation of motion of the mirror in the frame
   of reference S. \\

\noindent
SOLUTION 1: trivial manner \\

\vspace*{0.3cm}

Consider one photon (frequency $f'$) in the proper frame of the mirror.
Since the directions (and orientations) of the incident and outgoing photons are
characterized by
\begin{eqnarray}\label{1}
\vec{S} ' &=& \left ( + ~ \cos \theta ', \sin \theta ', 0 \right ) ~,
\nonumber \\
\vec{e} ' &=& \left ( - ~ \cos \theta ', \sin \theta ', 0  \right ) ~,
\end{eqnarray}
we can immediately write
\begin{eqnarray}\label{2}
p_{i}^{' \mu} &=& \frac{h~f'}{c} ~
		  \left ( 1, +~ \cos \theta ', \sin \theta ', 0 \right ) ~,
\nonumber \\
p_{o}^{' \mu} &=& \frac{h~f'}{c} ~
		  \left ( 1, -~ \cos \theta ', \sin \theta ', 0 \right ) ~,
\end{eqnarray}
for the four-momentum of the photon before interaction with the mirror
and after the interaction.

As a consequence, the mirror obtains four-momentum
\begin{equation}\label{3}
p^{' \mu} = p_{i}^{' \mu} ~-~ p_{o}^{' \mu} = \frac{h~f'}{c} ~
		  \left ( 0, 2~ \cos \theta ', 0, 0 \right ) ~.
\end{equation}

Application of the special Lorentz transformation to Eq. (3) yields
\begin{equation}\label{4}
p^{\mu} = \frac{h~f'}{c} ~2 ~ \gamma ~\left ( \cos \theta ' \right ) ~
	  \left ( \beta, 1, 0, 0 \right ) ~,
\end{equation}
where, as standardly abbreviated,
\begin{equation}\label{5}
\gamma = 1 ~/~ \sqrt{1 ~-~ \beta ^{2}}~~; ~~~\beta = v ~/~c ~.
\end{equation}

On the basis of Eq. (4), we can immediately write equation of motion
of the mirror
\begin{equation}\label{6}
\frac{d p^{\mu}}{d \tau} = \frac{E_{i}'}{c} ~ 2 ~ \gamma ~
			   \left ( \cos \theta ' \right ) ~
			   \left ( \beta, 1, 0, 0 \right ) ~,
\end{equation}
where $E_{i}'$ is the total energy (per unit time) of the incident radiation
measured in the proper frame of reference. \\

\vspace*{0.3cm}

\noindent
SOLUTION 2: application of general theory of Kla\v{c}ka (2000a) \\

\vspace*{0.3cm}

We have to choose orthonormal vectors in the systems S':
we will use $\vec{S} '$ and $\vec{e}_{1} '$ and one can easily find
\begin{eqnarray}\label{7}
\vec{S} ' &=& \left ( + ~ \cos \theta ', \sin \theta ', 0 \right ) ~,
\nonumber \\
\vec{e}_{1} ' &=& \left ( - ~ \sin \theta ', \cos \theta ', 0  \right ) ~.
\end{eqnarray}
We have to write ($Q_{2} ' = 0 $)
\begin{equation}\label{8}
\vec{p} ' =  \frac{h~f'}{c} ~ \left ( Q_{R}' ~\vec{S} '
	    ~+~ Q_{1} ' ~\vec{e}_{1} ' \right ) ~.
\end{equation}
On the basis of Eqs. (3), (7) and (8) we have
\begin{equation}\label{9}
Q_{R} ' = 2~ ( \cos \theta ' ) ^{2} ~~, ~~~
Q_{1} ' = -~2~ ( \sin \theta ') ~( \cos \theta ' )  ~; ~~~ ( Q_{2} ' = 0 ) ~.
\end{equation}
Other prescription yields
\begin{eqnarray}\label{10}
b_{i}^{0} &=& \gamma ~ ( 1 ~+~ \vec{v} \cdot \vec{S} ' / c ) =
	      \gamma ( 1 ~+~ \beta ~ \cos \theta ' ) ~,
\nonumber \\
\vec{b}_{i} &=& \vec{S} ' ~+~
	    \left [ ( \gamma ~-~ 1 ) ~ \vec{v} \cdot \vec{S} ' / \vec{v}^{2}
	    ~+~ \gamma / c \right ] ~ \vec{v} =
       \left ( \gamma ~\cos \theta ' ~+~ \gamma ~ \beta, \sin \theta ', 0 \right ) ~,
\end{eqnarray}
\begin{eqnarray}\label{11}
b_{1}^{0} &=& \gamma ~ ( 1 ~+~ \vec{v} \cdot \vec{e}_{1} ' / c ) =
	      \gamma ( 1 ~-~ \beta ~ \sin \theta ' ) ~,
\nonumber \\
\vec{b}_{1} &=& \vec{e}_{1} ' ~+~
	    \left [ ( \gamma ~-~ 1 ) ~ \vec{v} \cdot \vec{e}_{1} ' / \vec{v}^{2}
	    ~+~ \gamma / c \right ] ~ \vec{v} =
    \left ( -~ \gamma ~\sin \theta ' ~+~ \gamma ~ \beta, \cos \theta ', 0 \right ) ~.
\end{eqnarray}
Inserting Eqs. (9) -- (11) into Eq. (28) of Kla\v{c}ka (2000a), one obtains
\begin{eqnarray}\label{12}
\frac{d p^{\mu}}{d \tau} &=& \frac{E_{i}'}{c} ~\left \{
			     \left [ 2~ ( \cos \theta ' ) ^{2}  \right ] ~
			     \left ( b_{i}^{\mu} ~-~ \beta^{\mu} \right ) ~+~
			     \left [ -~2~ ( \sin \theta ') ~( \cos \theta ' )
			     \right ] ~ \left ( b_{1}^{\mu} ~-~ \beta^{\mu}
			     \right ) \right \}
\nonumber \\
			 &=& \frac{E_{i}'}{c} ~ 2 ~ \gamma ~
			     \left ( \cos \theta ' \right ) ~
			     \left ( \beta, 1, 0, 0 \right ) ~.
\end{eqnarray}

\vspace*{0.3cm}

\noindent
{\bf COMPARISON:} \\

Unit vectors $\vec{S} '$ and $\vec{e}_{1} '$ are used according to
Kla\v{c}ka (2000a). They lead to correct results.
The unit vectors are orthonormal in the system S'. How does the situation
look in the system S?
\begin{eqnarray}\label{13}
\vec{S} &=& \frac{1}{w '} ~ \left \{ \vec{S} ' ~+~
	    \left [ ( \gamma ~-~ 1 ) ~ \vec{v} \cdot \vec{S} ' / \vec{v}^{2}
	    ~+~ \gamma / c \right ] ~ \vec{v} \right \} ~,
\nonumber \\
w ' &=& \gamma ~ \left ( 1 ~+~ \vec{v} \cdot \vec{S} ' ~/~c \right ) ~,
\end{eqnarray}
and analogous equation holds for vector $\vec{e}_{1}$.
Inserting Eqs. (7), one obtains
\begin{eqnarray}\label{14}
\vec{S} &=& \left \{
	    \frac{\cos \theta ' ~+~ \beta}{1 ~+~ \beta ~ \cos \theta '},
	    \frac{\sin \theta '}{\gamma
	    \left ( 1 ~+~ \beta ~ \cos \theta '\right )}, 0 \right \} ~,
\nonumber \\
\vec{e}_{1} &=& \left \{
		\frac{-~\sin \theta ' ~+~ \beta}{1 ~-~ \beta ~ \sin \theta '},
	    \frac{\cos \theta '}{\gamma
	    \left ( 1 ~-~ \beta ~ \sin \theta ' \right )}, 0 \right \} ~.
\end{eqnarray}
It can be easily verified that scalar product of these two vectors is nonzero,
in general, even to the first order in $\beta$:
\begin{equation}\label{15}
\vec{S} \cdot \vec{e}_{1} \approx  \beta ~ ( \cos \theta ' ~-~ \sin \theta ' ) ~.
\end{equation}

\vspace*{1.0cm}

\noindent
SOLUTION 3: application of general theory of $G^{\mu ~\nu} ~ b_{i~ \nu}$
(Kla\v{c}ka 2001a).

\vspace*{0.3cm}

We will use equation of motion in the form:
\begin{equation}\label{16}
\frac{d \vec{p}'}{d \tau} = \frac{E_{i}'}{c}~ Q' ~ \vec{S} ' ~.
\end{equation}
Calculations yield
\begin{equation}\label{17}
Q' = Q_{i} ' ~-~ Q_{r} ' = diag(1, 1, 1) ~-~ diag(-~1, 1, 1) =
			   diag(2, 0, 0) ~.
\end{equation}
Moreover, expression for $\vec{S}$ presented in Eq. (14) yields
\begin{eqnarray}\label{18}
( Q' \vec{S} ) ^{T} &=& 2 ~ \left (
	 \frac{\cos \theta '~+~ \beta}{1 ~+~ \beta ~ \cos \theta '} ,
	 0, 0 \right )
\nonumber \\
( Q' \vec{\beta} ) ^{T} &=& 2 ~ \left ( \beta, 0, 0 \right ) ~,
\end{eqnarray}
\begin{eqnarray}\label{19}
\beta ^{T} ( Q' \vec{S} ) &=& 2 ~ \beta ~( \cos \theta ' ~+~ \beta ) ~/~
			    ( 1 ~+~ \beta ~ \cos \theta ' )
\nonumber \\
\beta ^{T} ( Q' \vec{\beta} ) &=& 2 ~ \beta^{2} ~.
\end{eqnarray}
Inserting into equation
\begin{equation}\label{20}
\frac{d p^{\mu}}{d \tau} = \frac{E_{i} '}{c} ~ G^{\mu ~ \nu} ~ b_{i~ \nu} ~,
\end{equation}
one obtains
\begin{equation}\label{21}
\frac{d p^{\mu}}{d \tau} = \frac{E_{i}'}{c} ~ 2 ~ \gamma ~
			   \left ( \cos \theta ' \right ) ~
			   \left ( \beta, 1, 0, 0 \right ) ~.
\end{equation}

\vspace*{1.0cm}

\noindent
REMARK: Inserting $E'_{i} = w^{2} ~S~ A' ~\cos \theta '$ into Eq. (6)
(or Eqs. (12), (21)) and using Eq. (14) for the purpose of obtaining
$\cos \theta ' = ( \cos \theta - \beta) / (1 - \beta ~\cos \theta)$, one
easily obtains: \\
i) $dE/d \tau = 2~\gamma^{3}~ S~A'~(\cos \theta - \beta )^{2} ~ \beta$;
using definition of radiation pressure $dE/dt \equiv P~v~A'$, we have
$P = 2~(S~/~c)~( \cos \theta - \beta)^{2} ~/~( 1 - \beta ^{2})$, or, \\
ii) $dp/d \tau = 2~\gamma^{3}~ (S~A'~/~c) ~(\cos \theta - \beta )^{2}$;
using definition of radiation pressure $P \equiv (dp/dt)~/~A'$, we have
$P = 2~(S~/~c)~( \cos \theta - \beta)^{2} ~/~( 1 - \beta ^{2})$.

Result for $P$ is consistent with result presented in Einstein (1905). \\

\subsection{Argument/statement 3}
{\it The assumption is that energy $E'$ of the particle is unchanged:
the energy of the incoming radiation equals the energy of the outgoing
radiation, per unit time. What this means is that the particle does not
heat up. Fine, as such, but wait. A physical explanation here involving
the Poynting-Robertson effect has the particle losing orbital energy due
to re-radiation effects, and so the size of its orbit reduces. So how does
it work, physically?}

{\it Answer:}

Yes, $dE'~/~d \tau = 0$ corresponds to conservation of mass of the particle.
However, the relation for energy holds in the proper reference frame of the particle
(rest frame of the particle), only. Lorentz transformation yields that
energy changes in other reference frames, e. g. in the rest frame of the
source of radiation. The special case $Q'_{1} = Q'_{2} =$ 0 in Kla\v{c}ka
(2000a -- e. g., Eqs. (7), (23)) -- corresponding to the P-R effect -- yields
that $dE / d t \ne$ 0 and the corresponding change of semi-major axis
is given by Eq. (14) in Kla\v{c}ka (1992b). \\

\subsection{Argument/statement 4}
{\it Where does the "really [realistically?] shaped particle come into it?"}

{\it Answer:}

Effective factors $Q'_{R}$, $Q'_{1}$, $Q'_{2}$ (see Eq. (7) in Kla\v{c}ka
2000a) take into account optical characteristics of the particle. They can
be calculated using, e. g., discrete dipole approximation -- see, e. g.,
Draine and Weingartner (1996). \\

\subsection{Argument/statement 5}
{\it $dE'~/~d \tau = 0$: If a momentum is transferred to the particle there
should be a corresponding gain in kinetic energy. In reality this term will
depend upon the particle's albedo etc.}

{\it Answer:}

The statement $dE'~/~d \tau = 0$ was discussed in section 2.3 -- conservation
of particle's mass. Even within a Newtonian physics we have $T = m ~v^{2}/2$,
$dT / dt = m ~\vec{v} \cdot \dot{\vec{v}}$ and this yields $dT' / dt =$ 0
in the proper frame of the particle, although $d \vec{p'} / dt \ne$ 0.
In reality $dE/dt \ne$ 0 in the rest frame of the source and $dE/dt$ depends
on optical properties of the particle (see sections 2.3 and 2.4).
(It has no sense to use the term "particle's albedo" if the size of the
particle is comparable to the wavelength of the incident electromagnetic
radiation, when measured in the proper frame of the particle.) \\

\subsection{Argument/statement 6}
{\it The problem is of great general interest and (at least at the Newtonian
level) has been treated intensively in the literature.}

{\it Answer:}

Since the time of Maxwell and Einstein (1905) we know that the discussed
problem cannot be treated at the Newtonian level. In reality, motion of
cosmic dust grains is standardly described by the P-R effect (of course,
it cannot be understood at the Newtonian level). P-R effect is a very
special case of the general equation of motion (Eq. (24) in
Kla\v{c}ka 2000a). \\

\subsection{Argument/statement 7}
{\it A real astronomical problem is oversimplified and the resulting equations
are not really useful.}

{\it Answer:}
Papers Kla\v{c}ka and Kocifaj (2001a, 2001b) show important difference between
orbital motion obtained on the basis of
Eq. (24) in Kla\v{c}ka (2000a) and the P-R effect. P-R effect
was used for six decades (up to now) and there was no astronomical protest that
"it is not really useful".

\section{Conclusion}
We have discussed some statements of astronomers as for more general
equation of motion than that given by the P-R effect. We have tried to
"stand up for" the general equation of motion, as we have presented it
for the first time in Eq. (24) in Kla\v{c}ka (2000a) and in a more
simple form in Kla\v{c}ka (2000a), in Kla\v{c}ka and Kocifaj (2001a).
We have shown that our more general equation of motion reduces
to the two cases discussed earlier in literature:
Einstein (1905), Robertson (1937). Any correct equation of motion has to be
consistent with the results of Einstein (1905) and Robertson (1937).

\acknowledgements{}
The paper was partially supported by VEGA grant No. 1/7067/20.

\end{document}